\documentclass[a4paper]{article}
\usepackage[utf8]{inputenc}
\usepackage{authblk}
\usepackage{lineno,hyperref}
\modulolinenumbers[5]
\usepackage{amsmath}
\usepackage{amsfonts}
\usepackage{breakcites}
\usepackage{graphicx, mathptmx}
\usepackage{caption}

\author[1]{A. Saha \thanks{arindamjal@gmail.com}}
\author[2]{S. Ghose\thanks{(Corresponding author: dr.souvikghose@gmail.com}}
\author[3]{A. Chanda\thanks{anirbanchanda93@gmail.com}}
\author[4]{B. C. Paul \thanks{bcpaul@associates.iucaa.in}}
\author[2]{Corresponding Author \thanks{dr.souvikghose@gmail.com}}
\affil[1]{Jalpaiguri Govt. Engineering College, Jalpaiguri, West Bengal, India 735102}
\affil[2]{HECRC \& ICARD, Physics Department,  University of North Bengal, Rajarammohunpur, West Bengal, India 734013}
\affil[3,4]{Physics Department,  University of North Bengal, Rajarammohunpur, West Bengal, India 734013}

\title{R\' enyi holographic dark energy in higher dimension Cosmology}
\date{\today}
\begin{document}
\maketitle
%%%%%%%%%%%%%%%%%%%%%%%%

\begin{abstract}
R\'enyi holographic dark energy (RHDE) correspondence is discussed in higher dimensional cosmology, namely  Kaluza-Klein (KK) cosmology. Both interacting and non-interacting cosmological scenario are considered here. It is found that the non-interacting model naturally leads to the late accelerated universe, unlike the standard Holographic Dark Energy models in 4$D$, which requires interaction to accommodate the late-time acceleration of the universe. The interacting model produces an accelerating universe at late time albeit failing to attain the estimated  present value of the deceleration parameter. The evolution of different cosmologically relevant parameters have been estimated. We consider here two  diagnostic tests namely, state-finder and $Om$ diagnostics to study the non-interacting model which is more favoured in the light of recent cosmological observations. Classical stability of the cosmological models are also discussed

\end{abstract}

\section{Introduction}
Present observations favour that our universe is flat \cite{spergel2003, komatsu2011} and it is  passing through a phase of accelerated expansion at the present epoch \cite{riess,perlm,perlm2,perlm3}. General Relativity (GR hereafter) can accommodate such a phase in the presence of a cosmological constant. However, the magnitude of the cosmological constant, as predicted by the present observations, can not be explained from any fundamental theory. Consequently, modifications of the gravitational sector or matter sector of the Einstein-Hilbert action are proposed in the literature to describe the present acceleration of the universe.  Recently, there has been a spurt in activities in the  modified theories of gravity, namely,  $f(R)$, $f(G)$, $f(T)$, (where $R$ represents the Ricci scalar, $G$ represents the Gauss-Bonnet terms and $T$ represents the torsion scalar) and other theories \cite{carroll, uddin, linderfg, samipaddy, gibb, nojde1, cappode, liucamp}. The modifications of GR assumes generic functions of the Ricci scalar or the Gauss-Bonnet terms or torsion scalar for the gravitational action instead of the linear term in  the Einstein-Hilbert action. The modified gravity theories posses  interesting properties that can resolve some of the important issues in cosmology including the accelerated expansion of the present universe. On the other hand, the modification of the matter sector can be considered in the presence of dynamical fluids with negative pressure unlike the cosmological constant. These are commonly known as Dark Energy (DE), and a volume of papers appeared in the literature discussing cosmological models with dynamical dark energy \cite{freese,overduin,fritzsch,basilakos,sola,ratra1988,gibb,armendariz2001,wetterich2017,escamilla2013}.  In this context thermodynamical scenario have been considered in the literature where the Benkenstein-Hawking (BH hereafter) entropy ( proportional to the area of the horizon) and the Holographic Principle (\textit{number of degrees of freedom of any physical system should scale with the boundary surface and not with the volume}) \cite{susskind, cohen, lii} play an important role.  The holographic principle essentially states that \textit{ in any physical system, the number of degrees of freedom should scale not with the volume but with the boundary surface}. Holographic dark energy  (HDE) models have been considered in the literature \cite{granda2008, gao2009, del2011, lepe2010, arevalo2014} to study cosmic problems. In the cosmological context, the choice of long-distance cutoff or infrared cut off (IR cut off hereafter) of HDE models is not uniquely defined. Hubble horizon is a natural candidate for IR cut off but as shown in \cite{hsu2004}, the choice fails to accommodate the late-time acceleration. Zimdahl and Pav\'on \cite{zimdahl2007} later shown that inclusion of an interaction could resolve such issues (Ref. \cite{miao2011} for a detailed review of different HDE models).
Klauza-Klein (KK hereafter) theory was originally studied with the hope of unifying gravity and gauge theories \cite{kaluza,klein} by introducing an extra dimension. KK theory has been used as a framework for many cosmological models to address some of the long-standing problems of the early and late universe which can not be addressed within GR. Some of these models adhere to the original motivation of KK theory with a compact extra dimension \cite{chodos1980has, mukhopadhyay2016dark} while others have considered non-compact theories where the extra dimension was treated as a parameter, defining 4$D$ universe embedded in a five-dimensional space-time \cite{wesson1984embedding}. The compactness, or the lack of it, is not the only source of variety in KK cosmology. Rich structure of the theory also stems from the presence of a source in 5$D$ in the presence and absence of cosmological constant, and inclusion of fewer or more degrees of freedom by the 5$D$ metric. The purest form of KK theory is worth studying if one is guided by the hope of the complete geometrization of physics as the theory is capable of inducing matter in $4D$ by $5D$ vacuum theory \cite{wesson1999space, paul1999exact}. The curvature of the five-dimensional space-time can also induce effective properties of matter in four dimensions (Campbell's theorem) \cite{wesson1993physical,wesson1996theory}. Another important property of classical KK theories is that they can be treated as emergent entropic gravity models \cite{verlinde2017emergent}. This makes the study of HDE models in KK gravity particularly interesting. In the present work, we propose a R\' enyi holographic dark energy correspondence in the compact framework of KK cosmology. 

The paper is organized as follows: in sec. (\ref{sec:reney} a brief overview of Re\'enyi entropy is presented, in sec. (\ref{sec:feqkk}) relevant field equations are developed, in sec. (\ref{sec:evun}), in sec. (\ref{sec:rsstable}) state finder diagnostics and stability of the model are studied, in sec. (\ref{sec:intc} an interacting RHDE model is discussed and finally in sec. (\ref{sec:disc}) results from interacting and non-interacting situations are compared and overall findings are summarized.
 
%*******************************************************************************************
\section{General idea of  R\' enyi entropy}
\label{sec:reney}

In non-extensive thermodynamics the Tsallis entropy ($S_T$) \cite{tsalis, biro2013q, czinner2016renyi, belin2013holographic} for a set of $W$ states is defined  as 
\begin{equation}
\label{tsaen}
S_{T}=k_{B} \frac{1- \sum_{i=1}^{W}p_{i}^{q}}{q-1} \; \; \left(\sum_{i=1}^{W}p_{i}=1; \; q\in \rm I\!R \right),
\end{equation}
where $p_i$ is the probability associated to the $i^{th}$ microstate with $\sum_{i}p_{i}=1$ and $q$ is any real number. Tsallis holographic dark energy has been extensively studied in the context of cosmology \cite{sharma2020diagnosing,varshney2019statefinder,thdemain} The original R\' enyi entropy ($S_{R}^{Org}$), on the other hand, is defined as \cite{tsalis}:
\begin{equation}
\label{rsaen}
S_{R}^{Org}=k_{B} \frac{ln\sum_{i=1}^{W}p_{i}^{q}}{q-1}=\frac{1}{1-q}ln\left[1+(1-q)S_{T}\right].
\end{equation}
It is interesting that both the eqs.(\ref{tsaen}) and (\ref{rsaen}) lead to Boltzmann-Gibbs entropy for $q=1$. Recently, it has been proposed that the Benkenstein-Hawking ($S_{BH}$) entropy too is a kind of non-extensive entropy which leads to a novel type of R\' enyi entropy \cite{czinner2016renyi} which is given by:
\begin{equation}
\label{rsaen1}
S_{R}=\frac{1}{\delta}ln\left(1+\delta S_{BH} \right),
\end{equation}
where $\delta=1-q$ and for $\delta=0$, $S_{R}=S_{BH}$.
In cosmology, R\'enyi holographics dark energy is extensively considered\cite{sharma2020statefinder,dubey2020diagnosing}. It is shown in \cite{moradpour2017accelerated} that when R\'enyi entropy is employed at the horizion it results in an accelerating universe in the Rastall framework. In the present work, the R\' enyi entropy is considered for describing the entropy on the Hubble horizon. As $S_{BH}=\frac{A}{4}$ (where $A$ is the area of the horizon), one obtains
\begin{equation}
\label{rsaen2}
S_{R}=\frac{1}{\delta}ln\left(1+ \frac{A}{4} \; \delta \right).
\end{equation}

%................................................................................................................

\section{Cosmology in KK framework and RHDE correspondence}
\label{sec:feqkk}
The Einstein field equation in higher dimension is given by:
\begin{equation}
\label{efeq}
R_{AB}-\frac{1}{2}g_{AB}R=\kappa T_{AB},
\end{equation}
where $A$ and $B$ runs from $0$ to $4$, $R_{AB}$ is the Ricci tensor, $R$ is the Ricci scalar and $T_{AB}$ is the energy-momentum tensor and $\kappa=8\pi G_{5}$ where $G_{5}$ is the five dimensional gravitational constant.
The $5$-dimensional space-time metric of the KK cosmology is (see Ref. \cite{ozel}) :
\begin{eqnarray}
\label{kkmetric}
ds^2 &=& dt^2-a^2(t)\left[\frac{dr^2}{1-kr^2}+r^2(d\theta^2 \right. \nonumber \\
&& \left. + sin^2\theta d\phi^2\right)+\left(1-kr^2)d\psi^2\right],
\end{eqnarray}
where $a(t)$ is the scale factor and $k=0,1(-1)$ represents the curvature parameter for flat and closed(open) universe. We consider a cosmological model where KK universe is filled with a perfect fluid.  The Einstein's field equation for the metric, given by (\ref{kkmetric}), becomes:
\begin{equation}
\label{kkfeq1}
\rho=6\frac{\dot{a}^2}{a^2}+6\frac{k}{a^2},
\end{equation}
\begin{equation}
\label{kkfeq2}
p=-3\frac{\ddot{a}}{a}-3\frac{\dot{a}^2}{a^2}-3\frac{k}{a^2}.
\end{equation}
For simplicity, we shall consider a flat universe ($k=0$). Using eq. (\ref{kkfeq1}) and eq. (\ref{kkfeq2})) we get for the time-time and space-space component of the fields equations as:
\begin{equation}
\label{kkfeq3}
\rho=6\frac{\dot{a}^2}{a^2},
\end{equation}
\begin{equation}
\label{kkfeq4}
p=-3\frac{\ddot{a}}{a}-3\frac{\dot{a}^2}{a^2}.
\end{equation}
The Hubble parameter is defined as $H=\frac{\dot{a}}{a}$ and from $T^{\mu\nu}_{;\nu}=0$ the continuity equation follows as:
\begin{equation}
\label{kkcont1}
\dot{\rho}+4H(\rho+p)=0.
\end{equation}
Using the  equation of state $p=\omega\rho$ in the equation of continuity in five dimension we get,
\begin{equation}
\label{kkcont2}
\dot{\rho}+4H\rho(1+\omega)=0.
\end{equation}
We now consider two types of cosmic fluids. Total energy density is expressed as $\rho=\rho_{D}+\rho_{m}$, where $\rho_{D}$ corresponds to the dark energy and $\rho_{m}$ is the matter energy density where $w_{m}=\frac{\rho_{m}}{p_{m}}$. For non-interacting fluid, the  conservation equations for ($p_{D},\rho_{D}$) and ($p_{m},\rho_{m}$) are separately satisfied:
\begin{equation}
\label{kkcontm}
\dot{\rho}_{m}+4H\rho_{m}(1+\omega_m)=0,
\end{equation}
\begin{equation}
\label{kkcontl}
\dot{\rho}_{D}+4H\rho_{D}(1+\omega_{D})=0.
\end{equation}
Let us define the dimensionless density parameters as:
\begin{equation}
\label{dpdef}
\Omega_m=\frac{\rho_m}{\rho_{cr}}, \; \;  \Omega_{D}=\frac{\rho_{D}}{\rho_{cr}},
\end{equation}
where $\rho_{cr}=6H^2$. Eq. (\ref{kkfeq3}) can be written in terms of density parameters as:
\begin{equation}
\label{denequ}
\Omega_m +\Omega_{D}=1.
\end{equation}
The ratio of the energy densities is given by:
\begin{equation}
\label{endenra}
r=\frac{\rho_m}{\rho_D}=\frac{\Omega_m}{\Omega_D}=\frac{1-\Omega_D}{\Omega_D}.
\end{equation}
Considering the time derivative of Eq. (\ref{kkfeq3}) and using Eqs. (\ref{kkcontm}), (\ref{kkcontl}), and (\ref{endenra}) we obtain:
\begin{equation}
\label{hubdot}
\frac{\dot{H}}{H^2}=-2\left(1+\omega_D \Omega_D \right)
\end{equation}

Blackhole thermodynamics can be implemented in describing the thermodynamics of cosmological models. In fact,  the first law of thermodynamics   $dE_T = TdS$ can be considered for the thermodynamical representation of the cosmological horizon. Here $E_T$ denotes the total energy of the universe, $T= \frac{1}{2\pi L}$,  the Cai-Kim temperature for a system with IR cutoff $L$, corresponds to the temperature of the de-Sitter horizon and apparent horizon in flat FRW cosmology, $S$ is the horizon entropy.  Similarly, in the DE dominated present universe, it is possible to take  the  relation $dE_D=\rho_D dV\propto dE_T = TdS$, where $E_D$ denotes the energy content for the DE representative.\\
 The apparent horizon is a proper casual boundary for a universe in agreement with the thermodynamical laws and thus the energy-momentum conservation law emerge out. Therefore, considering the Hubble radius as the IR cutoff ($L=H^{-1}$), we obtain the energy density for DE ($\rho_D$) candidate which is given by 
\begin{equation}
\label{eden}
\rho_D = B \; T\frac{dS}{dV},  
\end{equation}
where $B$ is the proportionality constant. Again using the temperature relation $T=\frac{H}{2\pi}$. The above expression can be integrated to get the functional form of $\rho_{D}$. RHDE correspondence can then be drawn following  Ref.\cite{moradpour2018},  in a 5-dimensional KK-cosmology,  we represent here the vacuum energy density which plays the role of the dark energy. We consider DE of the universe as given by RHDE in 5-dimension which is 
\begin{equation}
\label{rho-r}
\rho_{D}=\frac{B H^2}{8 \pi \left(1+\frac{\delta \pi}{H^2}\right)},
\end{equation}
where $B=3C^2$, and $C^2$ is a numerical constant. 

From Eq.(\ref{kkcontl}) and the time derivative of Eq.(\ref{rho-r}) one gets:
\begin{equation}
\label{rho-rdot}
\frac{\dot{\rho}_{D}}{\rho_{D}}=4\frac{\dot{H}}{H}\left(\frac{C^2-\Omega_D}{C^2}\right),
\end{equation}
where, 
\begin{equation}
\label{denpara-r}
\Omega_{D}=\frac{C^2}{2}\left(1+\frac{\delta \pi}{H^2}\right)^{-1}.
\end{equation}

%...............................................................................................................

\section{Evolution of universe}
\label{sec:evun}
Evolution of the universe, within the present model, can be explored with the field equations obtained in the previous section. The equation of state (EOS hereafter) parameter for the dark energy is obtained from eq. (\ref{kkcontl}), eq. (\ref{hubdot}), and eq. (\ref{rho-rdot}), which becomes:
\begin{equation}
\label{eosd-r}
\omega_{D}=\frac{C^2 - 2 \Omega_{D}}{C^2-2\Omega_{D} (C^2-\Omega_{D})}.
\end{equation}

The expression for $\Omega_D$, as obtained from eq. (\ref{hubdot}) and eq. (\ref{denpara-r}), is:

% Here I had to break the equation as it is a long equation and it can not be accommodated in a single column
\[
 \Omega_D=\frac{1}{2} \Big[ 1+2b(z+1)^4 \pm
\]
\begin{equation}
  \sqrt{1+4b(z+1)^4(1-C^2)+4b^2(z+1)^8} \Big],
  \label{eq:dden}
\end{equation}

where $b=\frac{\Omega_{D0}(1-\Omega_{D0})}{C^{2}-2\Omega_{D0}}$, with $\Omega_{D0}$ being the present value of the density parameter for dark energy.
%Figures are resized to fit the two column mode

\begin{figure}[htbp]
  \centering
  \includegraphics[scale=0.5]{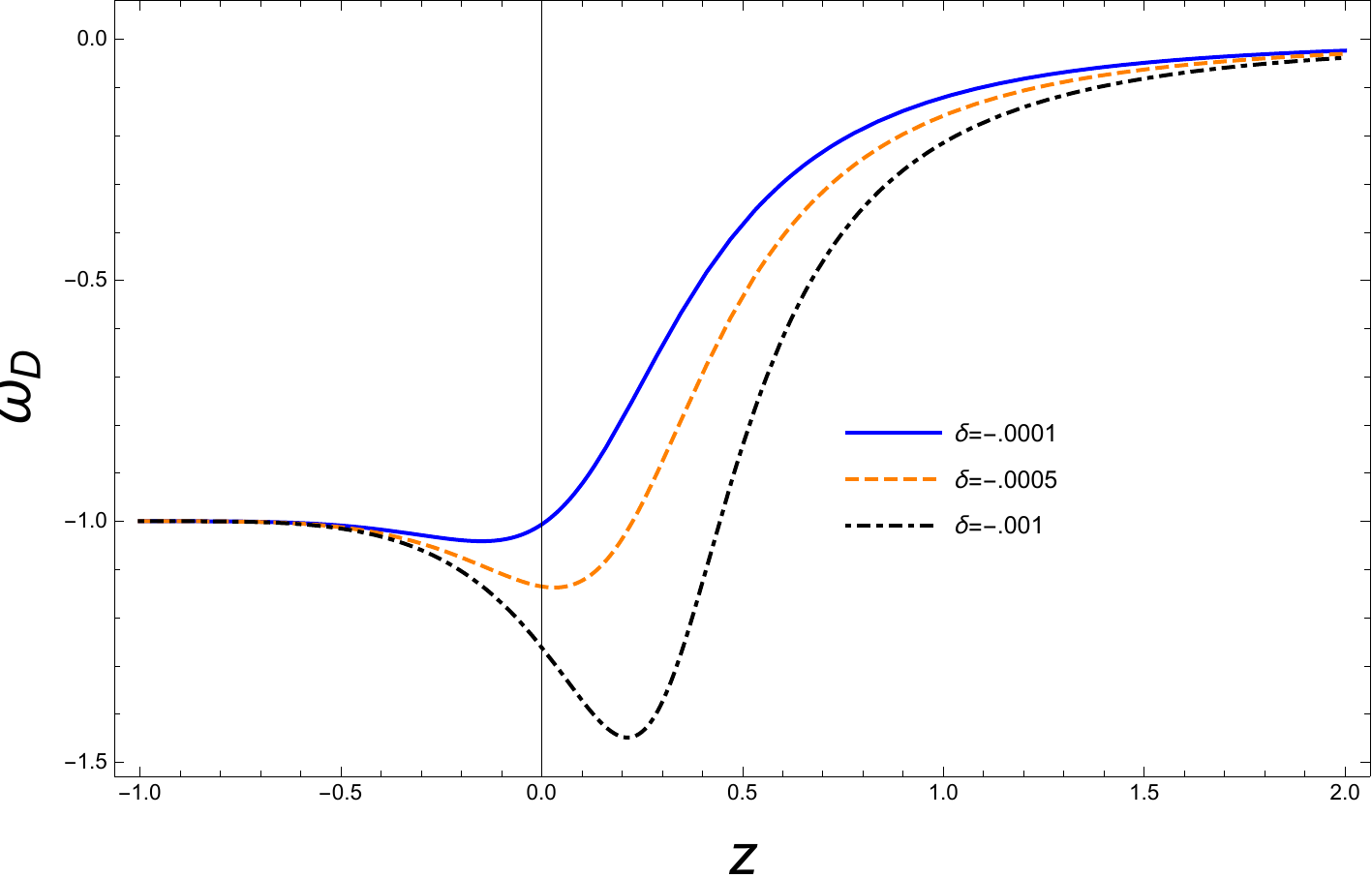}
  \caption{Evolution of dark energy EoS parameter $\omega_D$ for different values of model parameter $\delta$. A quintessence era in early universe followed by a phantom phase and finally convergence into another late time quintessence phase is seen}
  \label{fig:eos-z}
\end{figure}
The evolution of EoS parameter is shown in fig. (\ref{fig:eos-z}). The dark energy begins in a quintessence era ($\omega_D > -1$), but evolves into a phantom ($\omega_D < -1$) era, and finally converges to another quintessence era in late time. The initial and final phase of the EoS parameter is independent of the $\delta$ value, however it varies considerably with $\delta$ in the present epoch. For lower $\delta$ values the universe transits into the phantom era faster as compared to higher $\delta$ values. Another important parameter is the deceleration parameter ( $q=-\frac{\ddot{a}}{aH^2}$)which can be obtained from eq. (\ref{hubdot}) and eq. (\ref{eosd-r}):
\begin{equation}
\label{eq:decel}
q=\frac{C^2-2\Omega_{D}^2}{C^2-2\Omega_{D}(C^2-\Omega_{D})}
\end{equation}

\begin{figure}[t]
    \includegraphics[scale=0.5]{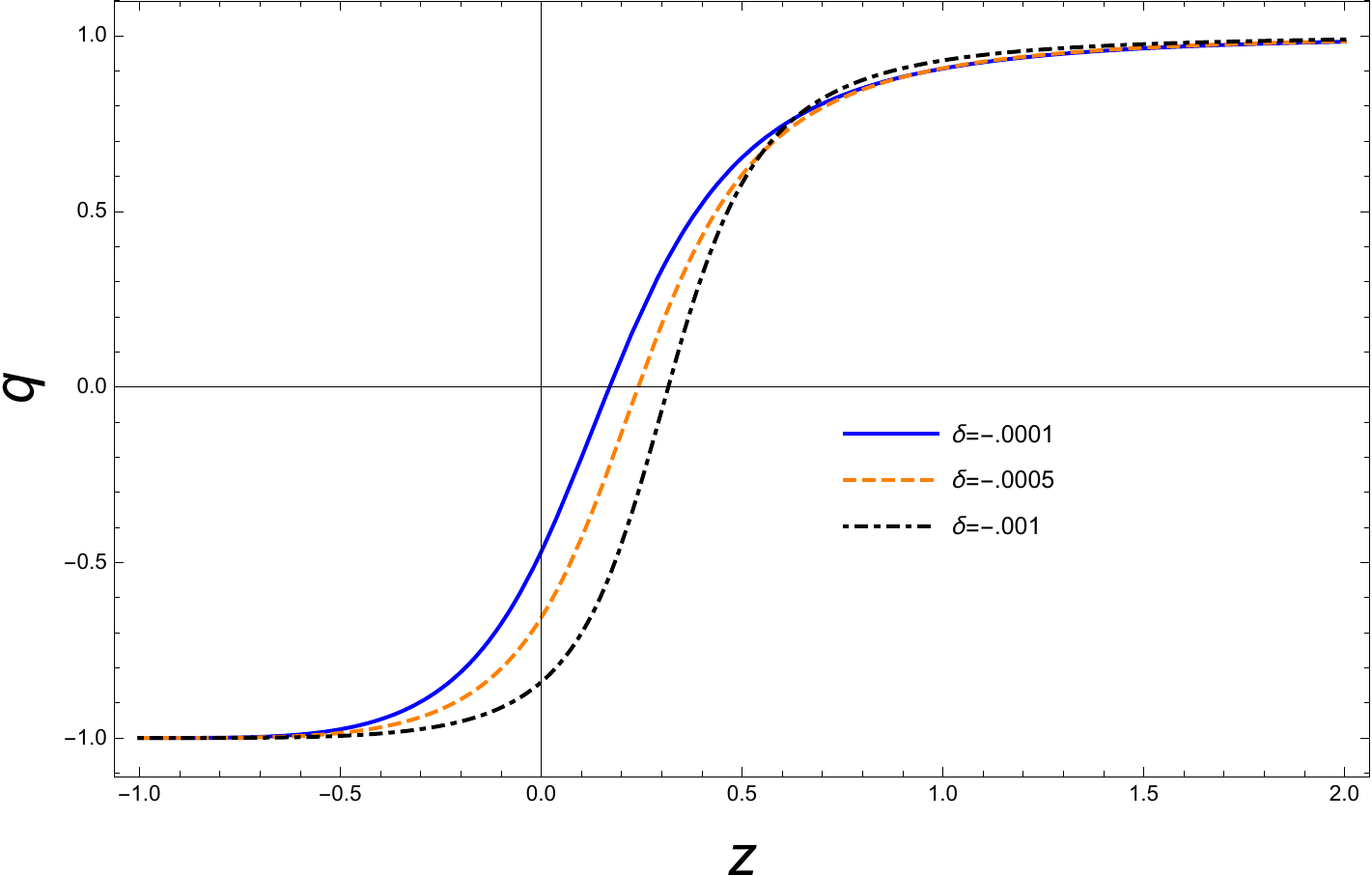}
  \caption{State of acceleration of universe at different stage of evolution. The deceleration parameter ($q$) value goes to negative in recent past, suggesting a late-time acceleration of universe. For different value of the model parameter $\delta$, present value of $q$ is seen to vary from $-0.5$ to $-0.1$ which is very much acceptable in light of supernovae data}
  \label{fig:qz}
\end{figure}
\begin{figure}[t]
  \includegraphics[scale=0.5]{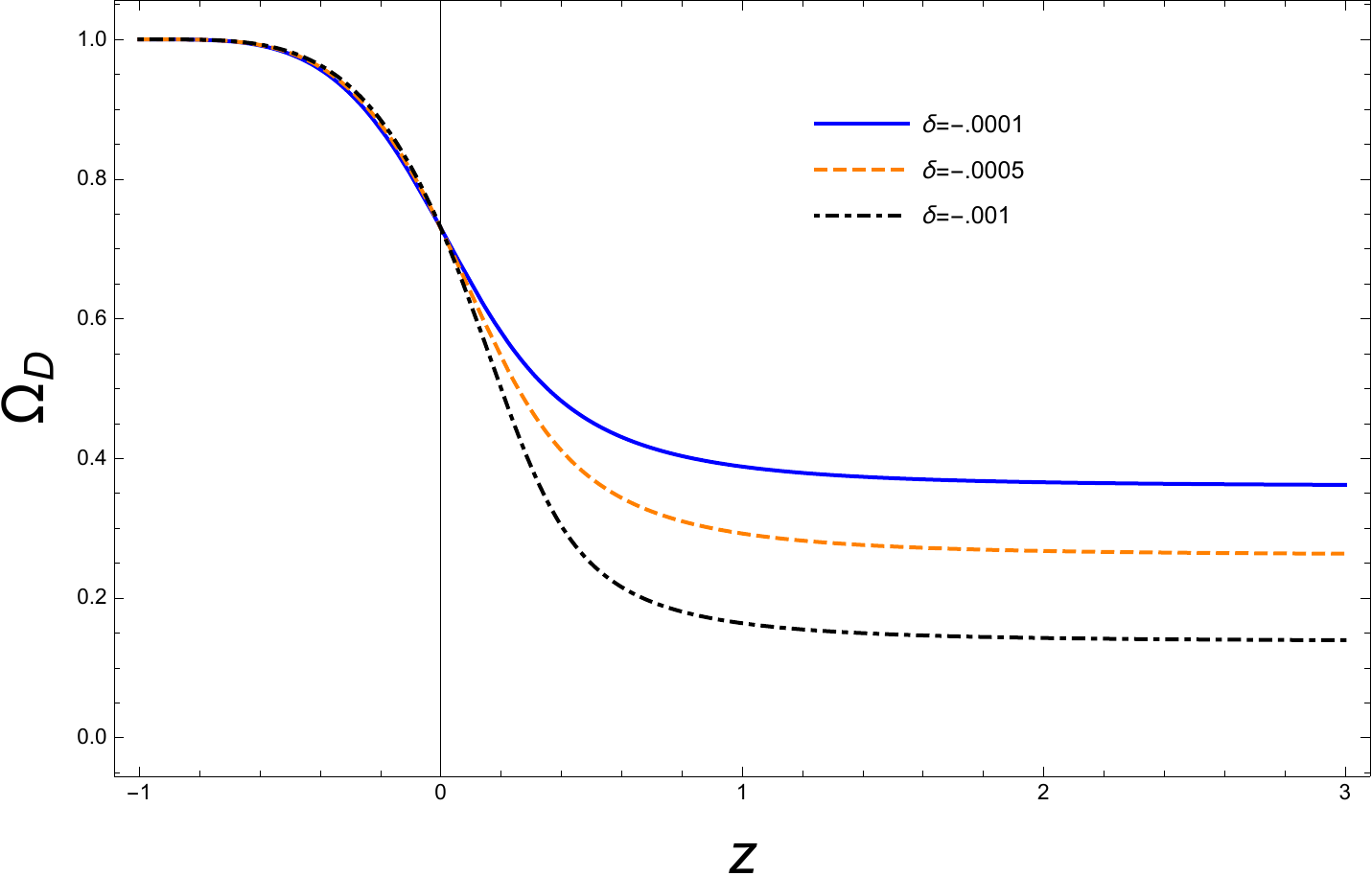}
  \caption{Evolution of density parameter $\Omega_D$ with redshift ($z$) for different values of model parameter $\delta$. The evolution converges to the present value of $\Omega_D\approx0.7$ which is acceptable. Dark energy clearly dominates the late-universe and gradually evolves to $\Omega_D=1$ in the future, regardless of $\delta$ value}
  \label{fig:omd-z}
\end{figure}

Fig. (\ref{fig:qz}) show the evolution of deceleration parameter. It is shown that the model permits late-time acceleration of the universe, and if $\delta$ is less the universe entered the accelerated phase earlier. Evolution of dark energy and its late-time dominance are evident from the evolution of density parameter $\Omega_D$ which is shown in fig. (\ref{fig:omd-z}). It is seen that the dark energy contribution was small earlier but at the present epoch it increases. It is also found that in future $\Omega_{D}$ attained a value which is independent of $\delta$, but in the past lower value of $\delta$ implies more $\Omega_{D}$.

%..................................................................................................................

\section{Diagnostics and stability}
\label{sec:rsstable} 
\begin{figure}
  \centering
  \includegraphics[scale=0.5]{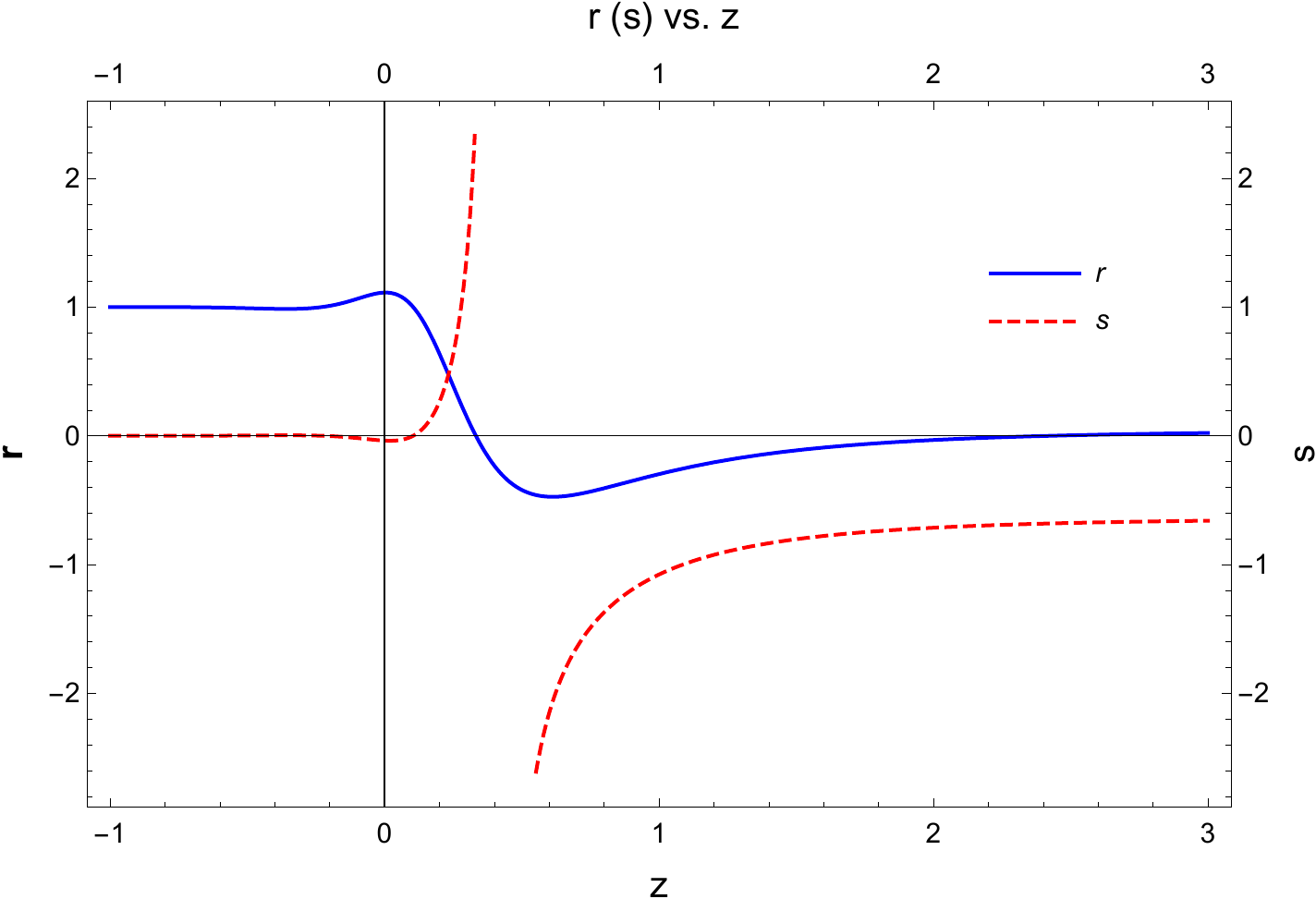}
  \caption{Plot of statefinder pairs with redishift ($z$). At present epoch the model is close to $\Lambda$CDM model. In near future the model coincides with $\Lambda$CDM model}
  \label{fig:rsz}
\end{figure}

The statefinder diagnostic pair (${r,s}$) was first introduced in ref. [\cite{sahni2003statefinder,alam2003exploring}]. The parameters are defined as:
\begin{equation}
  r\equiv\frac{\dddot{a}}{aH^{3}}, \; s\equiv\frac{r-1}{3(q-1/2)},
  \label{rsdef}
\end{equation}
where $q=-\frac{\ddot{a}}{aH^2}$ is the deceleration parameter. For the present model we study the evolution of $r(z)$ and $s(z)$ shown in fig. (\ref{fig:rsz}). For any $\Lambda$CDM model with non-zero $\Lambda$ the statefinder pair becomes (${1,0}$). In the present model the statefinder values very close to the $\Lambda$CDM model is obtained in the near future. So we conclude that the cosmological model admits a $\Lambda$CDM scenario in the negative redshift regime.

\begin{figure}[htbp]
  \centering
  \includegraphics[scale=0.5]{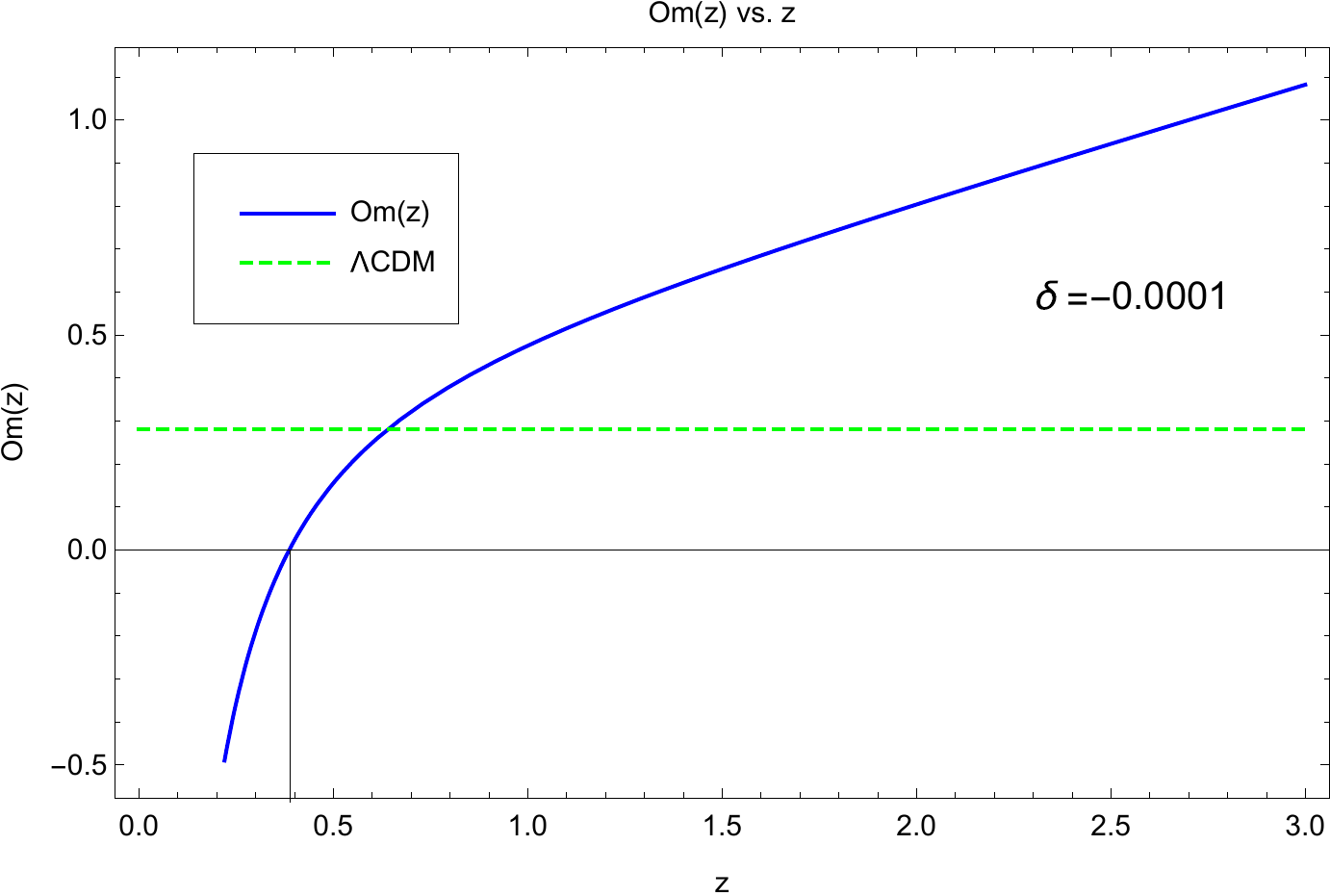}
  \caption{Evolution of $Om$ clearly suggests that the dark energy might have evolved from a phantom phase in recent past. Presently however it is of the quintessence nature. $Om$ is constant for the $\Lambda$CDM model and its value depends on $\Omega_{m0}$. A typical case of $\Lambda$CDM model is shown in the figure with for $\Omega_{m0}=0.27$. The blue solid line denotes evolution of the $Om$ parameter with redshift parameter $z$ and the green dashed line denotes the $\Lambda$CDM case}
  \label{fig:omz}
\end{figure}

To further explore the nature of dark energy we consider $Om$ diagnostic. To determine $Om$ diagnostic we define following (\cite{sahni2008two}):
\begin{equation}
  Om(x)=\frac{h^2(x)-1}{x^3-1}, \; x=1+z, \; h(x)=\frac{H(x)}{H_0}.
  \label{eqn:omdef}
\end{equation}
Comparison of the $Om$ values at two different points gives valuable insight about the nature of the dark energy. For example, if we consider two different $x$ values like $x_1$ and $x_2$ where $x_1<x_2$, for a $\Lambda$CDM model $Om(x_1,x_2)\equiv Om(x_1)-Om(x_2)=0$. If  $Om(x_1, x_2)<0$ the dark energy is phantom and for  $Om(x_1, x_2)>0$ the dark energy is of quintessence type. Evolution of $Om$ is shown in fig.(\ref{fig:omz}). The cosmological model with R\'enyi dark energy corresponds to quintessence matter at present. However it might recently made a transition from phantom phase (at around $z=0.39$).

Classical stability of a dark energy model can be discussed based on the squared sound speed $v_s^2=\left(\frac{\partial P}{\partial \rho}\right)$. For the present model, $v_s^2$ is given by:

\begin{equation}
v^{2}=\omega_{D}+\frac{(C^{2}-2\Omega_{D})\Omega_{D}}{(C^{2}-\Omega_{D})}\frac{\left[ C^2(C^2-2\Omega_{D})-(C^{2}-2\Omega_{D}^2)\right]}{\left[C^2-2\Omega_{D}\left(C^2-\Omega_{D}\right)\right]^2}.
\label{vsq}
\end{equation}
\begin{figure}[htbp]
  \centering
  \includegraphics[scale=0.5]{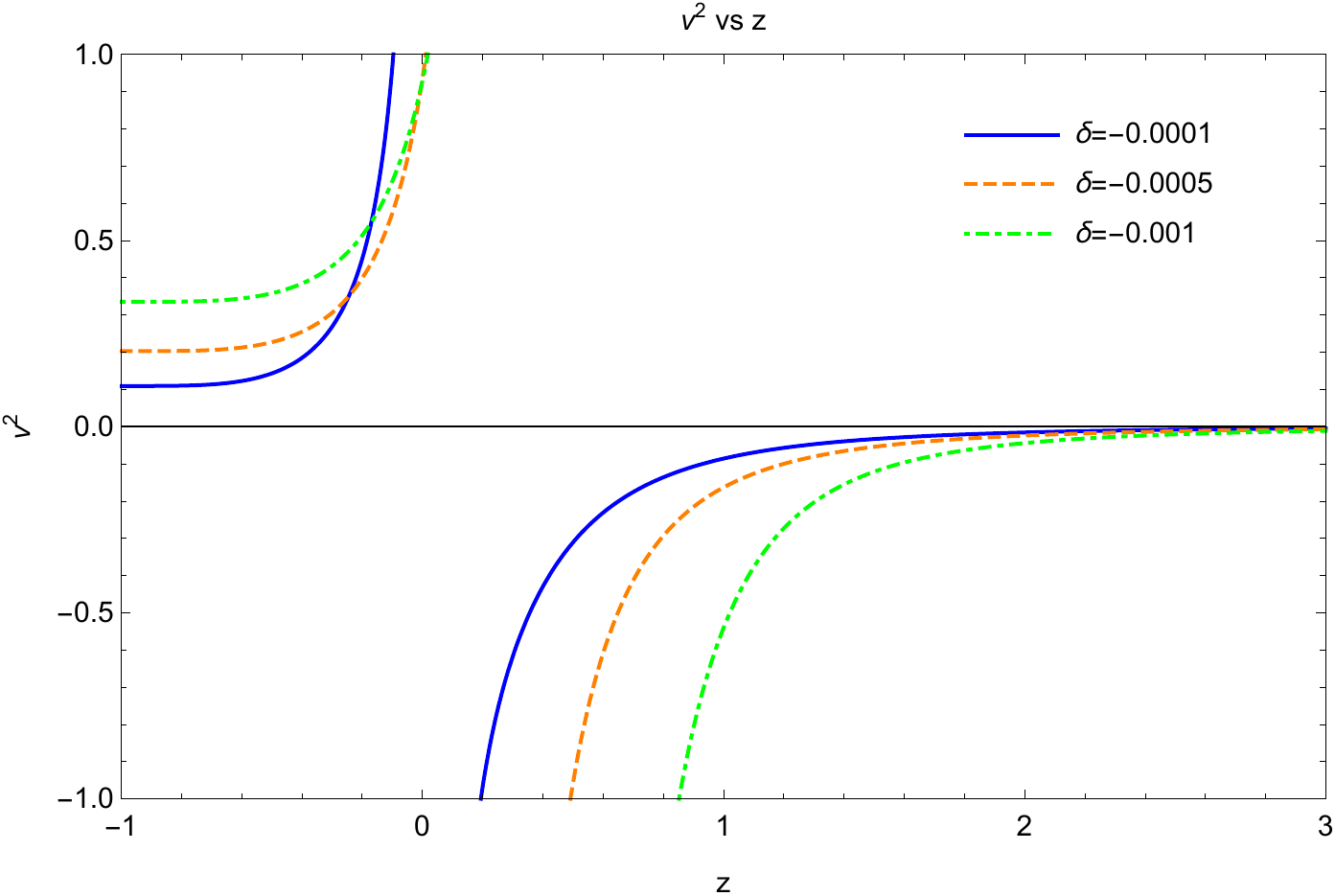}
  \caption{Classical stability of the non-interacting RHDE model}
  \label{fig:vsqz}
\end{figure}
 The evolution of the squared sound speed is shown in fig. (\ref{fig:vsqz}). The cosmological model remains unstable ($v^2<0$) in the past. However, the model is stable ($v^2>0$) at the present epoch and will remain so in future.

 \begin{table}
  
        \centering
    \begin{tabular}{|c|c|c|}
      \hline 
      Parameter & Observation & Non Interacting \\ 
			      &  & Model \\ 
      \hline 
      $q_0$ & $\approx -0.6 \pm 0.02$ & $\approx -0.6$ \\ 
      \hline 
      $\omega_{D0}$ & $\approx -1$ & -1.013 \\ 
      \hline 
      $\Omega_{D0}$ & $0.6911 \pm  0.0062$ & 0.7329 \\ 
      \hline 
    \end{tabular}
    \caption{Cosmological Parameters for non-interacting RHDE model.} 
		 \label{tab:values}
  \end{table}
%-------------------------------------------------------------------
\section{Interacting RHDE model}
\label{sec:intc}
Considering two types of fluids that are interacting. We study the evolution of the universe here. The total energy density is then given by: $\rho=\rho_{D}+\rho_{m}$, where $\rho_{D}$ is density of the dark energy  and $\rho_{m}$ that of the matter. The conservation equations for ($p_{D},\rho_{D}$) and ($p_{m},\rho_{m}$) are separately satisfied for the non-interacting fluids. For the interacting dark energy models, however, we have:
\begin{equation}
\label{ikkcontm}
\dot{\rho}_m+4H\rho_{m}(1+\omega_m)=Q,
\end{equation}
\begin{equation}
\label{ikkcontl}
\dot{\rho}_{D}+4H\rho_{D}(1+\omega_{D})=-Q,
\end{equation}
where $Q$ gives the interaction between the dark energy and the dark matter.
%-------------------------
\begin{figure}[htbp]
  \centering
  \includegraphics[scale=0.5]{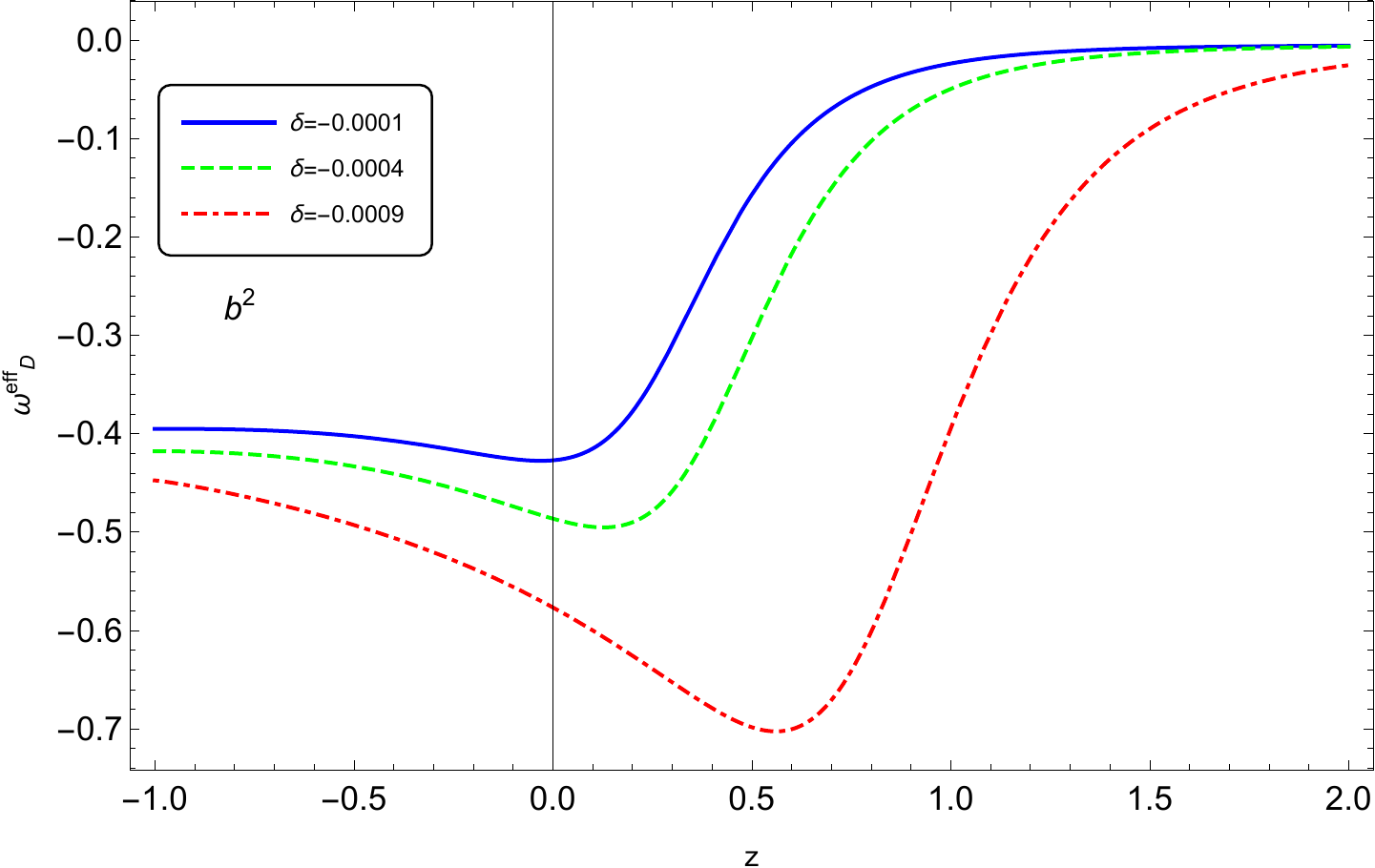}
  \caption{Evolution of EOS parameter with redshift for different values of  $\delta$}
  \label{fig:iomzd}
\end{figure}
%-----------------------------
As suggested in \cite{sharifkk}, let us consider $Q=\Gamma \rho_{D}$ and denote the ratio of the energy densities with $r$ ($r=\frac{\rho_m} {\rho_{D}}$); $\Gamma$ being the decay rate.  An effective equation of state can be defined (see \cite{setarehol}) as:
\begin{equation}
\label{eq:efeosdef}
\omega^{eff}_{D}=\omega_{D}+\frac{\Gamma}{4H} \; \; and \; \; \omega^{eff}_{m}=-\frac{1}{r} \frac{\Gamma}{4H}.
\end{equation}
The continuity equations as obtained from above consideration are:
\begin{equation}
\label{kkcontm2}
\dot{\rho}_m+4H\rho_m(1+\omega^{eff}_m)=0,
\end{equation}
\begin{equation}
\label{kkcontl2}
\dot{\rho}_{D}+4H\rho_D(1+\omega^{eff}_{D})=0.
\end{equation}
Eq. (\ref{eq:efeosdef}), eq. (\ref{rho-rdot}), and (\ref{hubdot}) lead to an expression for the effective equation of state parameter:
\begin{equation}
\label{eq:wfef2}
\omega_{D}^{eff}=\frac{(C^2-2\Omega_{d}) - 2b^2(C^2-\Omega_{d})}{C^2-2\Omega_{d}(C^2-\Omega_{d})}.
\end{equation}
%-----------------------
\begin{figure}[htbp]
  \centering
  \includegraphics[scale=0.5]{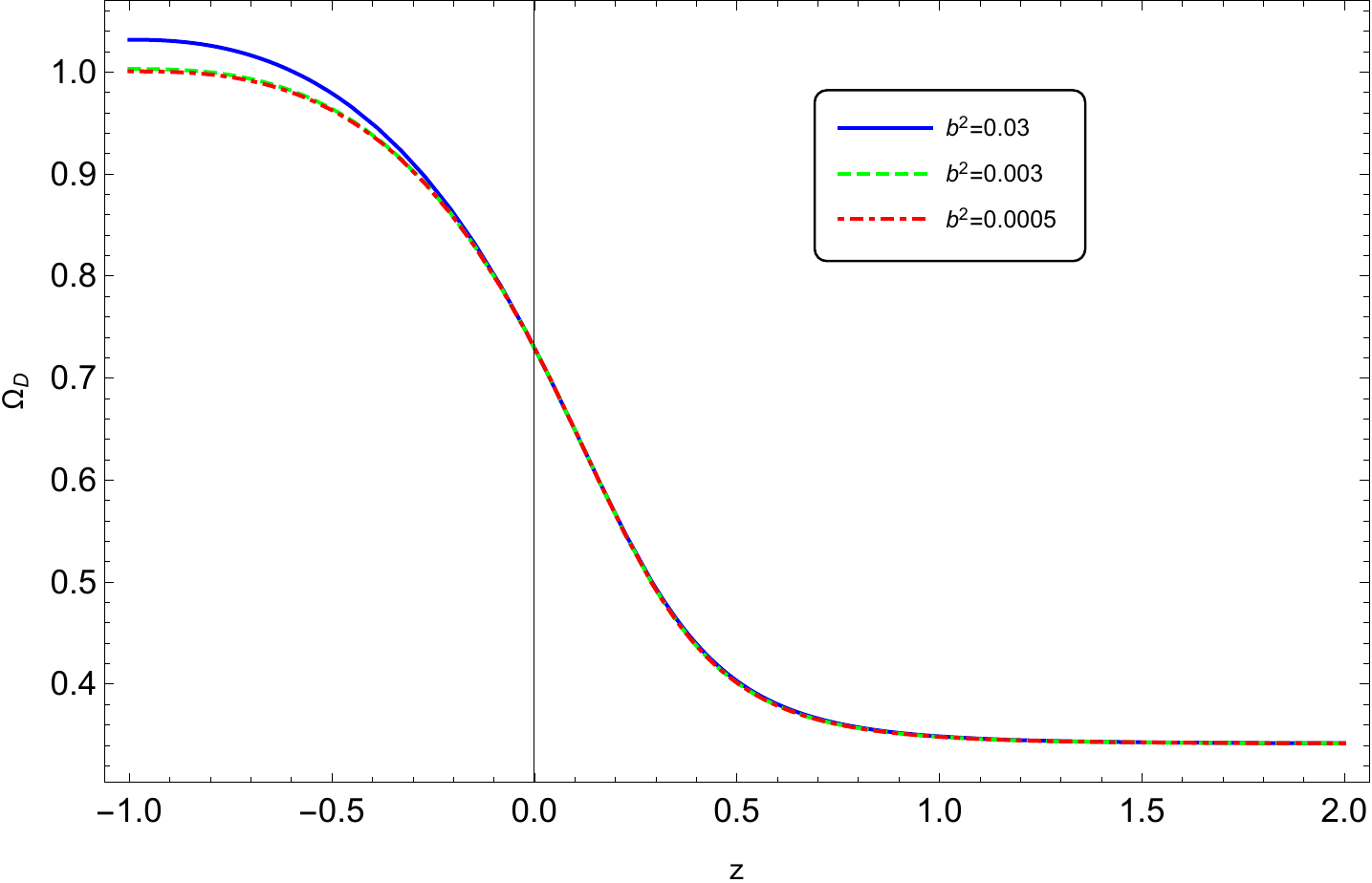}
  \caption{Evolution of density parameter with redshift for different values of  coupling constant}
  \label{fig:icomz}
\end{figure}
\begin{figure}[htbp]
  \centering
  \includegraphics[scale=0.5]{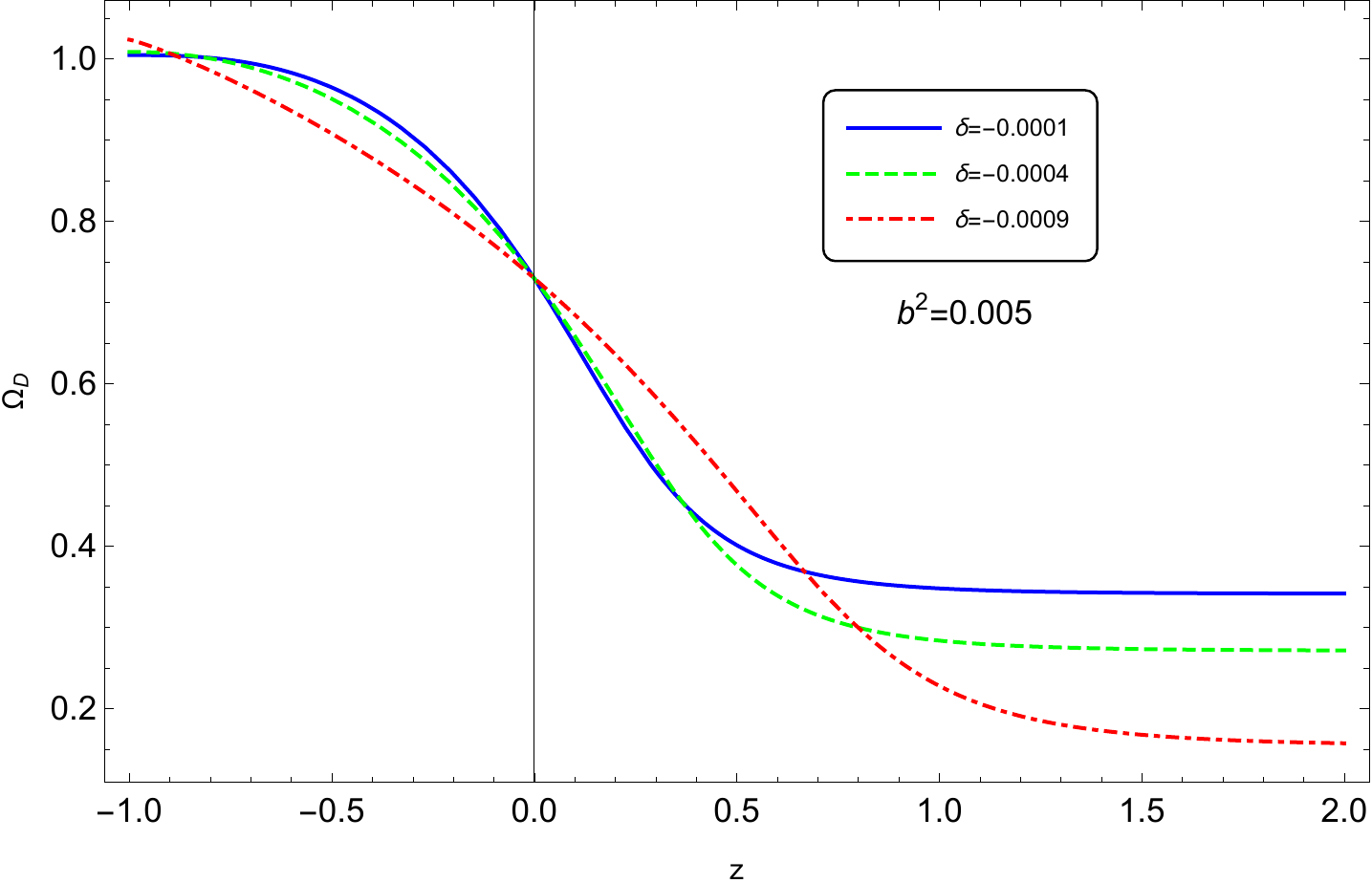}
  \caption{Evolution of density parameter with redshift for different values of  $\delta$}
  \label{fig:icomd}
\end{figure}
%-----------------------------------------------------
Evolution of the effective EOS parameter ($\omega_{D}^{eff}$) is shown in fig. (\ref{fig:iomzd}). The effective EOS parameter fails to achieve the observationally favoured value which is close to $-1$. Eq.(\ref{denpara-r}) and eq. (\ref{hubdot}) gives:
\begin{equation}
\label{eq:omprim}
\Omega^{'}_{D}=\frac{4\Omega_D(c^2-2\Omega_D)\left[c^2(1-\Omega_D)-b^2 \Omega_D (c^2-\Omega_D)\right]}{c^2-2\Omega_D(c^2-\Omega_D)}.
\end{equation}
where $\Omega_{D}'=-(1+z)\frac{d\Omega_{D}}{dz}$.
As seen from the eq. (eq:omprim) evoulution of the density parameter depends on both the model parameter $\delta$ and the coupling parameter $b^2$. It is seen that the future evolution of the density parameter is more affected by the variation in the coupling parameter value (fig. (\ref{fig:icomd}). For example, although the dark energy dominates in the future universe, its share is suppressed for a lower value of $b^2$ as seen from Fig.(8). On the other hand, the past evolution of the density parameter is more dependent on the model parameter ($\delta$) which is plotted in fig. (\ref{fig:icomd}. For a lower value of $\delta$ indicates that the share of dark energy was significantly lower in the past than that with a higher value of $\delta$. In the future the epoch the dependence on $\delta$ will reverse. The deceleration parameter ($q$) is obtained using eq. (\ref{hubdot}) and eq. (\ref{eq:wfef2}).
%-------------------------
\begin{figure}[htbp]
  \centering
  \includegraphics[scale=0.5]{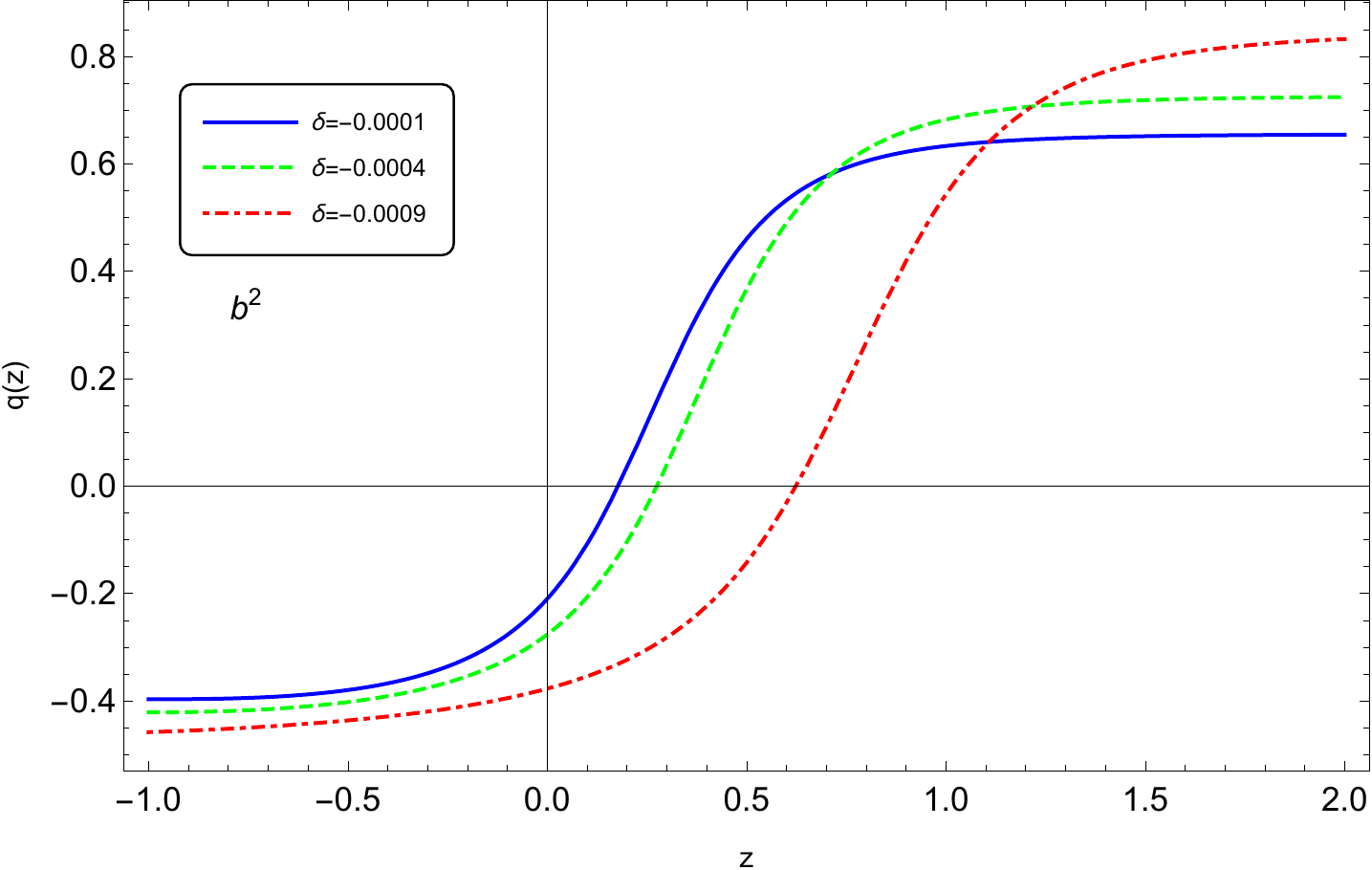}
  \caption{Evolution of deceleration parameter with redshift in interacting model}
  \label{fig:iqzd}
\end{figure}
%------------------------------
\begin{equation}
\label{eq:iqz}
q=\frac{c^2-2\Omega_{D}^2-4b^2\Omega_{D}(c^2-\Omega_{D})}{c^2-2\Omega_D(c^2-\Omega_D)}.
\end{equation}
Evolution of deceleration parameter with redshift $z$ is shown in the fig. (\ref{fig:iqzd}). The transition redshift is different for different values of $\delta$. More importantly, it is seen that the deceleration parameter value goes below $-0.4$ while the observationally favoured value for the present epoch is around $-0.6$ which maybe attained for a less value of $\delta$. Finally, classical stability of the model can be investigated calculating the adiabatic sound speed, which is given by:
\begin{equation}
\label{eq:ivsq}
v^2=\omega_{d}^{eff}+\frac{M}{(c^2-\Omega_{D})\big[c^2-2\Omega_{D}(c^2-\Omega_{D})\big]^2},
\end{equation}
where,
\begin{equation*}
\begin{aligned}
M={} & (c^2-2\Omega_{D})\Omega_{D}\Big[\Big.-c^2(2b^2-1)(c^2-2\Omega_{D}) \\
     & +(b^2-1)(c^2-2\Omega_{D}^2)\Big. \Big].
\end{aligned}
\end{equation*}
Evolution of the classical stability is shown in fig. (\ref{fig:ivsq}). We note that the KK-model, is not stable. This is in contrast with the interacting fluids with standard holographic dark energy models and Tsallis holographic dark energy models, studied in the compact KK framework, where a classically stable cosmological model can be obtained (for detail see. \cite{ushol, usthde}). In an interacting cosmological model we do not find a classically stable universe and it does not yield present value of the cosmological parameters in the range suggested by the observations. Thus, an interacting cosmological model with R\'enyi dark energy in a 5-dimensional KK-model is not acceptable at the present epoch. However, the interaction might have occurred  at the early universe. On the other hand, for a non-interacting RHDE with other fluids the cosmological model obtained here is found stable at the present epoch but it permits an unstable universe in  the past.
%----------------------------
\begin{figure}[htbp]
  \centering
  \includegraphics[scale=0.5]{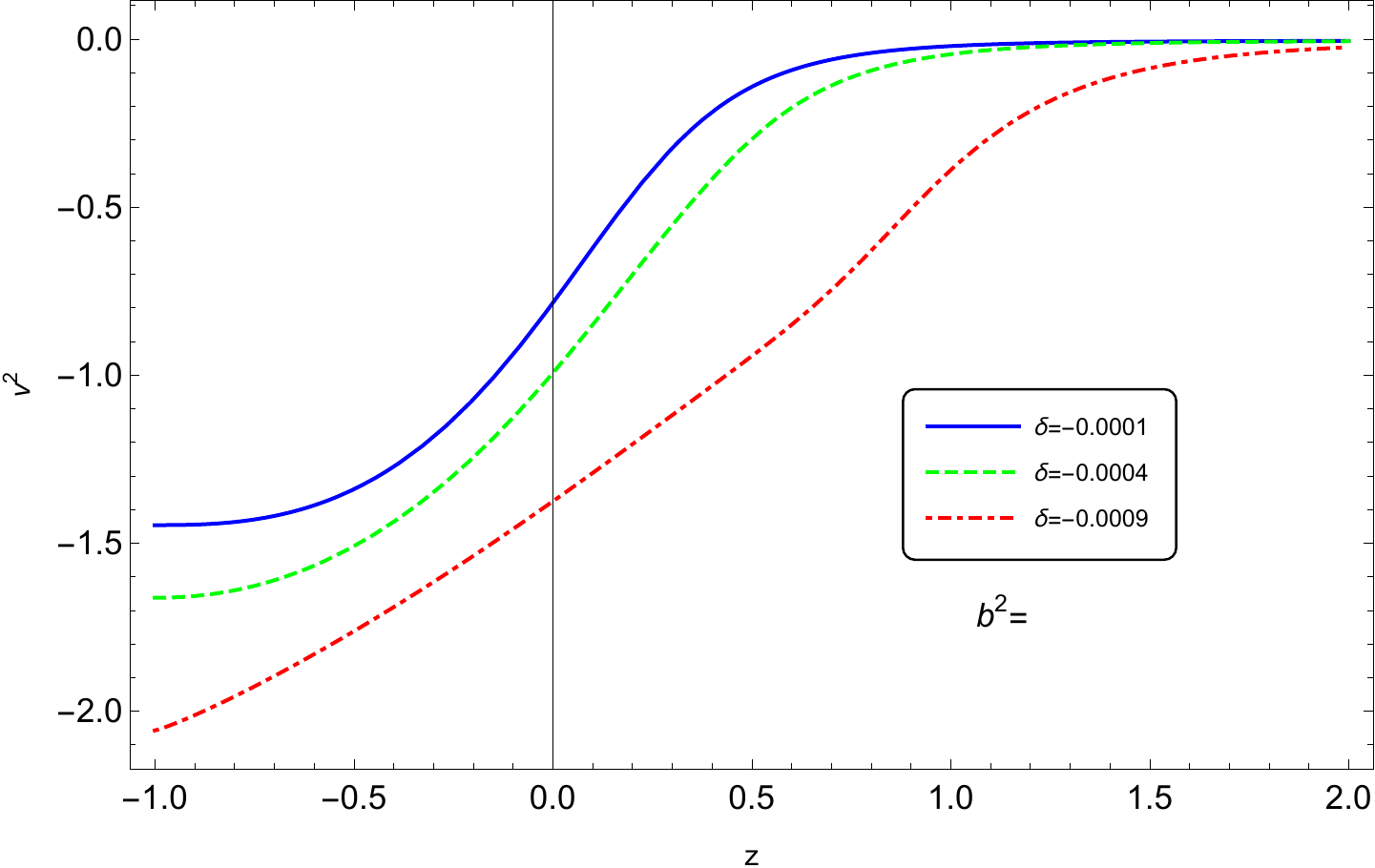}
  \caption{Classical stability of the interacting RHDE model}
  \label{fig:ivsq}
\end{figure}
%--------------------------------------------------------------------

  \begin{table}
  
        \centering
    \begin{tabular}{|c|c|c|}
      \hline 
       Parameter & Observation& Interacting \\ 
			  & & Model \\ 
      \hline 
      $q_0$ & $\approx -0.6 \pm 0.02$ & $\approx -0.4$ \\ 
      \hline 
      $\omega_{D0}$ & $\approx -1$ & -0.4 \\ 
      \hline 
      $\Omega_{D0}$ & $0.6911 \pm  0.0062$ & 0.7300 \\ 
      \hline 
    \end{tabular}
    \caption{Cosmological Parameters for interacting RHDE model} 
		 \label{tab:ivalues}
  \end{table}

%.............................................................................................................

\section{Discussion}
\label{sec:disc}

The holographic dark energy (HDE) models are worth discussing particularly in the absence of a definite quantum theory of gravity. KK framework was originally proposed with the hope of unifying gravity with the other gauge theories. Consequently, the study of HDE models in the KK framework is interesting in itself. HDE models in cosmology are explored in the KK gravity [\cite{sharifkk,ushol}]. Recently, Tsallis and R\'enyi holographic dark energy models are considered in cosmology as a non extensive extension of HDE.  In the present work, both non-interacting and interacting  R\'enyi holographic dark energy models are explored in the KK gravity framework.  As mentioned earlier that accommodating a late time phase of acceleration is often challenging in HDE models without interaction in the dark sector. We show here that with the R\'enyi  entropy in KK framework, the late-time acceleration of the universe can be naturally realised. Table (\ref{tab:values}) shows a comparison between observationally favored values of different cosmological parameters and the same obtained from the present model. Interestingly, whereas in the case of standard holographic dark energy models in $4D$, interaction is needed for viable late time behavior, R\'enyi holographic dark energy in KK can achieve this feat without interaction between dark energy and dark matter. It is seen that the interacting model fails to produce observationally accepted values of the cosmological parameters at the present epoch. While the non-interacting model has a prolonged classically stable region and the non-interacting model remains classically stable at every epoch. It was shown in \cite{symt} that classically stable non-interacting R\'enyi holographic dark energy can be obtained in loop quantum gravity also. The instability of the interacting model also may be specific to the prescription discussed here to introduce the interaction in the first place.  Note that the prescription works well with standard HDE [\cite{sharifkk,ushol}].  It is interesting that in \cite{bhat} the author also found the R\'enyi model which is classically unstable. There, it was noted that the sound speed was highly sensitive to the model parameter $\delta$. It was also observed that the negativity of sound-speed became more pronounced for a lower value of $\delta$. In the present work the finding has been in the same tune as is apparent from fig. (\ref{fig:ivsq}). The difference between Benkenstein and R\'enyi entropy has been noted earlier in \cite{ ghaffari2020inflation}, where the authors also showed how R\'enyi  entropy would successfully describe a late-time acceleration as well as an early inflationary phase.  Present observations suggest that the dark energy EoS parameter is very close to $-1$. The interacting model cannot attain such a phase in any stage of evolution. In the non-interacting model, on the other hand, the dark energy EoS parameter is very close to $-1$ in the present epoch and accumulates $\Lambda$CDM in the future. In course of evolution non-interacting dark energy passes through an initial quintessence phase, a phantom phase, and finally another quintessence phase. The interacting dark energy model, however, never passes through any phantom phase. Two different diagnostic tests, namely Statefinder and Om diagnostics have been done with the non-interacting model. These tests independently confirms that in the future the dark energy model will coincide with  a $\Lambda$CDM cosmology. The interacting model considered here has not been investigated further in absence of any classically stable region at any epoch. In both the scenarios considered here, cosmological parameters are sensitive in near future and near past on the model parameters $\delta$, and in the case of the interacting model, they are sensitive to decay rate too. Thus it is found that a  flat universe described by non-interacting R\'enyi  dark energy in KK-model is favourable for a stable present universe.
Present observations may put a reasonable constraint on the model parameters, but that is beyond the scope of the present discussion and would be taken up in a separate work.

\section*{Acknowledgements}
The authors would like to thank IUCAA Centre for Astronomy Research \& Development (ICARD), North Bengal University  for extending research facilities. AC would like to thank University of North Bengal for providing senior research fellowship. BCP would like to thank DST-SERB Govt. of India (File No.: EMR/2016/005734) for a project support. SG would like to than HECRC, NBU for visiting associateship.

%\section*{References}

%\nocite{*}

%\bibliography{rhderef}

\end{document}